\documentclass[aps,prl,preprint,groupedaddress]{revtex4-1}
\usepackage{graphicx}
\begin{document}

\noindent\textbf{A WAY TO DISCOVER MAXWELL'S EQUATIONS\\ THEORETICALLY}\\

\date{}

\hspace{2.7cm}
\begin{minipage}[t]{110mm}
\textbf{Krzysztof R\c ebilas}\\

\emph{Zak\l{}ad Fizyki. Akademia Rolnicza im. Hugona Ko\l{}\l{}\c ataja. \\
Al. Mickiewicza~21, \mbox{31-120} Krakow. Poland.\\
 Telephone: \mbox{48-12-6624390}. Fax: \mbox{48-12-6336245}.\\
E-mail: krebilas@ar.krakow.pl}
\end{minipage}\\
\vspace{1.5cm}

\noindent The Coulomb force, established in the rest frame of a source-charge $Q$, when transformed to a new frame moving with a velocity $\vec{V}$ has a form $\vec{F}= q\vec{{E}} + q\vec{v} \times \vec{{B}}$, where  
 $\vec{{E}}=\vec{E}'_\parallel + \gamma\vec{E}'_\perp
$ and $\vec{{B}}= \vec{\frac{V}{c^2}} \times \vec{{E}}$ and $\vec{E}'$ is the electric field in the rest frame of the source. The quantities $\vec{E}$ and $\vec{B}$ are then manifestly interdependent. We prove that they are determined by Maxwell's equations, so they represent the electric and magnetic fields in the new frame and the force 
 $\vec{F}$ is the well  known from experiments Lorentz force. In this way Maxwell's equations may be discovered theoretically for this
 particular situation of uniformly moving sources. The general solutions of the discovered Maxwell's equations lead us to fields produced 
by accelerating sources. \\


\section*{1. INTRODUCTION}
Maxwell's equations,
\begin{equation}
\vec{\nabla} \cdot  \vec{E} = \frac{\rho}{\epsilon_0} 
\end{equation}
\begin{equation}
\vec{\nabla} \times \vec{B} -\frac{1}{c^2}\frac{\partial \vec{E}}{\partial t}=\frac{\vec{j}}{c^2 \epsilon_0}
 \end{equation}
\begin{equation}
\vec{\nabla} \times \vec{E} + \frac{\partial \vec{B}}{\partial t} =0
\end{equation}
\begin{equation}
\vec{\nabla} \cdot \vec{B} =0
\end{equation}
and the Lorentz force law 
\begin{equation}
\vec{F}= q\vec{{E}} + q\vec{v} \times \vec{{B}},
\end{equation}
are generalizations based on many different experiments on the forces between electric charges and currents. The Lorentz force law is
usually introduced as an assumption separate from the field equations. Nevertheless the entire structure appears to be Lorentz covariant. Since the theory of relativity establishes some relationships among Maxwell's equations and between electric and magnetic fields, this suggest that the equations of classical electrodynamics could be derived from fewer experimental data by using relativistic arguments.

 In this paper we want to reveal such a possibility.
It is our desire to introduce Maxwell's equations in a manner which more clearly exhibits the role of special relativity. Namely, we show that to find out Maxwell's equations it is enough to know from experiments only Coulomb's law (established in the rest frame of the source-charge).  Using the Lorentz transformation of coordinates we are able to
determine the force and its components (the fields $\vec{E}$ and $\vec{B}$) produced by the source-charge when it is in motion with constant velocity. We prove that defined within this formalism fields $\vec{E}$ and $\vec{B}$ undergo Maxwell's equations. In this way the experimental way of discovering Maxwell's equations may be replaced to some extend by a theoretical study.

Certainly, our method cannot be applied to accelerating  sources because we do not know the Lorentz transformation between accelerating frames. Surprisingly Maxwell's equations we derive for the special case, i.e.  charges in uniform motion, have also much more general solutions depending  on the acceleration of sources.  The discovered theoretically laws lead us then to the completely general electromagnetic fields. Nevertheless one must be aware that while the physical correctness of the solutions of Maxwell's equations for uniformly moving charges is \emph{logically} guarantied  by our reasoning (provided our space-time structure is Minkowskian), validity of  the general solutions requires experimental verification. 

Let us explain it in more detail in the context of the known views on the issue.
It may seem that the results presented in this paper are in contradiction to the opinion of Feynman. 
In his lectures he wrote that the statement that "all of electrodynamics can be deduced solely from Lorentz
transformation  and Coulomb's law (...) is completely false."[1] Next he explains that to obtain Maxwell's \emph{fields} one 
must assume the retardation of interactions and that the electromagnetic potentials do not depend on acceleration
 of source. The textbook by W. G. V. Rosser [2] is the most famous work in which the development of Maxwell's fields is
 performed  on the basis of these assumptions. The equivalent set of assumptions is used by Frisch and Wilets [3] 
in their paper. If concerns our work, the standpoint we represent is essentially different. First of all, we do not 
derive directly Maxwell's fields but Maxwell's \emph{equations}. Next we argue that the general Maxwell's fields can 
be obtained simply as the \emph{solutions} of the discovered equations.  Certainly, in such an attitude no assumptions 
of those ones listed by Feynman must be separately postulated; simply they are embodied in the general solutions of
 Maxwell's laws. It follows then that, contrary to the Feynman point of view, the proposed in this paper way to Maxwell's
 equations, and next to Maxwell's fields, requires only knowledge of Coulomb's law and the Lorentz transformation. 

But this is not the end of the story. In a sense Feynman is right. Namely, although one can derive Maxwell's equations
 from merely  Coulomb's law and the Lorentz transformation, the general solutions of the derived Maxwell equations are 
\emph{not equivalent} to the assumed laws. The final result, i.e. fields $\vec{E}$ and $\vec{B}$ depending on the 
acceleration of source, exceeds the situation of uniformly moving sources we consider to derive Maxwell's equations
 and "suggests" also that the most general force law is the Lorentz one instead of Coulomb's law
 (even if the source is at rest, a nonzero acceleration-dependent magnetic field is produced). But let us emphasize: it is
 only a "suggestion". While the fields $\vec{E}$ and $\vec{B}$ for sources moving with constant velocity are well 
defined through the Lorentz transformation of Coulomb's law, the acceleration dependent electric and magnetic fields
 are completely different and cannot be expressed by means of Coulomb's force. In the context of our reasoning the 
latter ones occur as a bonus but not necessary physically correct. This is why they need to be confirmed by additional 
experiments. 

Concluding, actually "all of electrodynamics"  does not amount to Coulomb's law and the Lorentz transformation but at 
the same time we show in this paper that the Maxwell theory can be theoretically \emph{discovered} (to be later verified 
experimentally) merely on the basis of these laws.  

The presented here approach at some points may resemble the Rosser [2] method of developing Maxwell's equation for
 uniformly moving sources. However Rosser derives each Maxwell's equation in a different manner. Gauss' law is verified 
by means of integration of the explicit expression for $\vec{E}$. Faraday's law is checked by integration and 
differentiation of the explicit formulas on $\vec{E}$ and $\vec{B}$. In turn, Gauss' law for magnetism is proved
 on the basis of a field-line picture. The Ampere-Maxwell law is developed using purely vector analysis but its 
inhomogeneous form is derived only for the macroscopic ("averaged") charge distributions and fields. In contrast 
to it our method is purely algebraic and the same for all Maxwell's equations. In this way one may clearly see the 
role of Coulomb's law and the Lorentz transformation occurring on the background of the vector identities, which is 
not so apparent in the method of Rosser. It seems also that mathematically our method is more concise and strict,
 especially if one compares the way how the transformation of forces is performed and how the important relations 
between the time and spatial derivatives are introduced  (Rosser usses a pictorial method).

One can find an axiomatic development of the laws of electrodynamics in the work of Lehmberg $[4]$. At crucial points, 
however, the author performs generalization of Gauss' law to a Lorentz-covariant law, which is a weak point of theoretical 
argumentations since generalizing means simply guessing. Similar methods are presented in the papers of Krefetz $[5]$ and 
Kobe $[6]$. 
 Maxwell's \emph{fields} are also derived by Tessman $[7]$ who uses Coulomb's law and a set of several additional 
assumptions (among them is Newton's third law for steady state charge distributions). In the textbook by Purcell [8]
 electric and magnetic fields are developed for a special configuration of steady currents using the force transformation. 
A heuristic way to introduce Maxwell's equations is shown in the textbook of M. Schwartz~$[9]$.  We believe that our
 deductive method which does not require neither extraordinary  postulates nor any generalizations to get Maxwell's 
equations   is a nice exemplification of a power of theoretical reasoning based solely on the properties of the Lorentz 
transformations. The generalizations we need to justify entirely Maxwell's theory are \emph{not} required
 \emph{for obtaining} Maxwell's equations, as it is done in the cited papers, but refer only to the validity
 of the general solutions of the independently discovered laws. \\

\section*{2. DEFINITIONS}
Assume in a system of reference $S'$ a force $\vec{F}'$ is exerted on a body. From
 the point of view of some observer placed in an another frame $S$ in the  same situation the measured force is $\vec{F}$. If the 
frame $S'$ has a velocity $\vec{V}$  with respect to the system $S$, then the relation between the forces is (see Appendix):
\begin{equation}
\vec{F}=  \vec{F}'_\parallel + \gamma\vec{F}'_\perp +  \gamma\vec{v} \times \left(\vec{\frac{V}{c^2}} \times \vec{F}' \right),
\label{FFF}
\end{equation}
where $\gamma=(1- V^2/c^2)^{-1/2}$,  $\vec{v}$ is a velocity of the body measured in the system $S$ and the indices $\parallel$ and  $\perp$ refer to the directions parallel and perpendicular to the velocity $\vec{V}$. The force $\vec{F}$ may be rewritten in the following form:
\begin{equation}
\vec{F}= \vec{\mathcal{E}} + \vec{v} \times \vec{\mathcal{B}},
\label{EB}
\end{equation}
where:
\begin{equation}
\vec{\mathcal{E}}=\vec{F}'_\parallel + \gamma\vec{F}'_\perp
\label{EO}
\end{equation}
and
\begin{equation}
\vec{\mathcal{B}}= \gamma \left(\vec{\frac{V}{c^2}} \times \vec{F}' \right).
\end{equation}
Note however that we can write the definition of $\vec{\mathcal{B}}$ also as follows:
\begin{equation}
\vec{\mathcal{B}}= \gamma \left(\vec{\frac{V}{c^2}} \times \vec{F}' _\perp\right) = \gamma \left(\vec{\frac{V}{c^2}} \times \frac{\vec{\mathcal{E}} _\perp}{\gamma}\right),
\end{equation}
and finally get:
\begin{equation}
 \vec{\mathcal{B}}= \vec{\frac{V}{c^2}} \times \vec{\mathcal{E}}.
\label{BO}
\end{equation} 
The equation (\ref{EB}) with definitions (\ref{EO}) and (\ref{BO}) are quite general because we have not introduced any restrictions on the force $\vec{F}'$. 

Consider now interaction between two charged point particles. Let a charge $Q$ be a source of force $\vec{F}'$   that acts on a test charge
 $q$. Let the charge $Q$ be at rest in some inertial system $S'$. The force $\vec{F}'$  acting on the particle $q$ at position $\vec{r}\;'$ is given by the Coulomb law:
\begin{equation}
\vec{F}'=\frac{qQ}{4\pi \epsilon_0}\frac{\vec{r}\;'-\vec{r}\;'_Q}{|\vec{r}\;'-\vec{r}\;'_Q|^3},
\label{COULOMB}
\end{equation}
where $\vec{r}\;'_Q$ it is the position of the charge $Q$ in the frame $S'$. 
If we express this force by means of electric field
 intensity $\vec{E}'$
\begin{equation}
 \vec{F}'=q\vec{E}',
\label{F'E'}
\end{equation}
the Coulomb law may be written in the equivalent form of Gauss' law as:
\begin{equation}
\vec{\nabla}' \cdot  \vec{E}' = \frac{\rho'(\vec{r}\;')}{\epsilon_0},
 \label{GAUS'}
\end{equation}
where $\rho'(\vec{r}\;')=Q\delta(\vec{r}\;'-\vec{r}\;'_Q)$.  Eq. (\ref{COULOMB}) shows clearly that the force $\vec{F}'$ (so also $\vec{E}'$) is independent of the velocity of the test charge $q$. Also the Coulomb force (and the field $\vec{E}'$) in the frame $S'$ is static in the sense that it does not depend explicitly on time $t'$.

Using Eq. (\ref{F'E'})  and definitions (\ref{EO}) and (\ref{BO}) we can now rewrite Eq. (\ref{EB}) as follows:
\begin{equation}
\vec{F}= q\vec{{E}} + q\vec{v} \times \vec{{B}},
\label{EB1}
\end{equation}
where
\begin{equation}
\vec{{E}}=\vec{E}'_\parallel + \gamma\vec{E}'_\perp
\label{E}
\end{equation}
and
\begin{equation}
\vec{{B}}= \vec{\frac{V}{c^2}} \times \vec{{E}}.
\label{B}
\end{equation} 
In our reasoning the charges $q$ and $Q$ are treated as scaling factors invariant under the Lorentz transformation. 
This assumption follows from the experimental data which show that charge is a velocity-independent quantity.

Because $\vec{E}'$ does not depend on the test charge velocity,  from the definitions (\ref{E}) and (\ref{B})  follows that the fields $\vec{E}$ and $\vec{B}$ as functions of $\vec{E}'$
also do not depend on the test particle velocity.
The force given by Eq. (\ref{EB1}) separates then on the term that does not depend on the charge $q$ velocity $\vec{v}$, i.e. 
$q\vec{E}$ and the term velocity-dependent: $q\vec{v} \times \vec{{B}}$ determining the force perpendicular to the velocity of moving particle. 
  Eq. (\ref{EB1}) seems then to represent the Lorentz force law. To prove it is true we must show that the vectors $\vec{E}$ (which we will call the electric field intensity in the frame $S$)
 and $\vec{B}$ (which defines the magnetic field induction in the frame $S$)  satisfy Maxwell's equations. Validity
 of the Coulomb law in  the rest frame of the source charge $Q$  (which we will use in the form of the Gauss' law (\ref{GAUS'}))  is the 
only assumption which is needed to perform this task.\\

\section*{3. GAUSS' LAW FOR ELECTRICITY: $\vec{\nabla} \cdot  \vec{E} = \frac{\rho}{\epsilon_0}$}
Let us calculate the divergence  $\vec{\nabla} \cdot  \vec{E}$ for the field $\vec{E}$ in an arbitrary system of reference $S$.
Since the operator $\vec{\nabla}$ transforms as a vector under the spatial rotation of  coordinates, the divergence  $\vec{\nabla} \cdot  \vec{E}$ is a scalar function which value does not depend on the orientation of the axes of the system of reference. So,  we choose the $x-$axis along the velocity $\vec{V}$ and apply the Lorentz transformation in its simple form: $x'=\gamma (x-V t)$, $y'=y$, $z=z'$. From the relation (\ref{E}) we have the components of electric field $\vec{E}$:
\begin{equation}
E_x=E_x' \qquad E_y=\gamma E_y' \qquad  E_z=\gamma E_z'
\end{equation}
Thus:
\begin{equation} 
\vec{\nabla} \cdot  \vec{E}= \frac{\partial E_x}{\partial x} + \frac{\partial E_y}{\partial y} + \frac{\partial E_z}{\partial z}
 = \frac{\partial x'}{\partial x}\frac{\partial E_x'}{\partial x'}+ \gamma \frac{\partial E_y'}{\partial y'} + \gamma \frac{\partial E_z'}{\partial z'} = 
\gamma \vec{\nabla}' \cdot  \vec{E}'= \gamma \frac{1}{\epsilon_0}Q\delta(\vec{r}\;'-\vec{r}\;'_Q),
\label{gaus}
\end{equation}
where we have dropped the term $\frac{\partial t'}{\partial x}\frac{\partial E_x'}{\partial t'}$ because ${ E_x'}$ does not depend on $t'$.
But:
\begin{eqnarray}
\delta(\vec{r}\;'-\vec{r}\;'_Q) = \delta(x'-x_Q')\delta(y'-y_Q')\delta(z'-z_Q')= \delta(\gamma(x-x_Q))\delta(y-y_Q)\delta(z-z_Q)=\nonumber \\
=\frac{1}{\gamma}\delta(x-x_Q)\delta(y-y_Q)\delta(z-z_Q)=\frac{1}{\gamma} \delta (\vec{r}-\vec{r}_Q),
\label{ddd}
\end{eqnarray}
where we have used the identity:
\begin{equation}
\delta(f(x))=\sum_n\frac{\delta(x-x_n)}{|f'(x_n)|}
\end{equation}
Inserting the result  Eq. (\ref{ddd}) into Eq. (\ref{gaus}) we conclude that the field $\vec{E}$ measured in the arbitrary frame $S$ satisfies the Gauss' law:
\begin{equation}
\vec{\nabla} \cdot  \vec{E} = \frac{1}{\epsilon_0}Q \delta (\vec{r}-\vec{r}_Q)
\label{I}
\end{equation}
because $ Q \delta (\vec{r}-\vec{r}_Q)$ is simply the charge density $\rho(\vec{r})$ of a point particle $Q$ in the frame $S$.

Note that intead of using the Dirac-delta function one may get the same result by considering the density $\rho'$ to be defined
 as $\rho'=Q/(dx'dy'dz')$. Next, due to the Lorentz contraction $dx'=\gamma dx$ we get $\rho'=Q/(\gamma dxdydz)=\rho/\gamma$, which corresponds
 to the relation given in Eq. (\ref{ddd}).\\

\section*{4. AMPERE-MAXWELL LAW: $\vec{\nabla} \times \vec{B} -\frac{1}{c^2}\frac{\partial \vec{E}}{\partial t}=\frac{\vec{j}}{c^2 \epsilon_0} $}
Using  the following vector identity:
\begin{equation}
\vec{\nabla} \times (\vec{C} \times \vec{D} )= \vec{C}(\vec{\nabla} \cdot \vec{D}) - \vec{D}(\vec{\nabla} \cdot \vec{C}) + (\vec{D} \cdot \vec{\nabla})\vec{C} -
(\vec{C} \cdot \vec{\nabla})\vec{D}
\label{ATB}
\end{equation}
we have for the field $\vec{B}$:
\begin{equation}
\vec{\nabla} \times \vec{B}= \vec{\nabla} \times (\frac{\vec{V}}{c^2} \times \vec{E} )= \frac{1}{c^2}\vec{V}(\vec{\nabla} \cdot \vec{E}) - \frac{1}{c^2}\vec{E}(\vec{\nabla} \cdot \vec{V}) + \frac{1}{c^2}(\vec{E} \cdot \vec{\nabla})\vec{V} -
\frac{1}{c^2}(\vec{V} \cdot \vec{\nabla})\vec{E}
\label{21}
\end{equation}
Because $\vec{V}$ is a constant parameter of the Lorentz transformation (formally it may be treated as a constant "field") hence the second and the third term on the right hand side of the last equation vanish. So we get:
\begin{equation}
\vec{\nabla} \times \vec{B}= \frac{1}{c^2}\vec{V}(\vec{\nabla} \cdot \vec{E}) - 
\frac{1}{c^2}( \vec{V} \cdot \vec{\nabla} )\vec{E}.
\label{22}
\end{equation}
First we find the equivalent form for the operator  $\vec{V} \cdot \vec{\nabla}$. The Lorentz transformation of spatial coordinates for an arbitrary 
direction of the velocity $\vec{V}$ is [10]:
\begin{equation}
\vec{r}\;'=\vec{r} + \frac{\gamma-1}{V^2}(\vec{V} \cdot \vec{r})\vec{V} - \gamma \vec{V}t,
\label{LOR1}
\end{equation}
and for the time coordinate:
\begin{equation}
t'=\gamma \left(t- \frac{\vec{V} \cdot \vec{r}}{c^2}\right).
\label{LOR2}
\end{equation}
It follows that:
\begin{equation}
\frac{\partial }{\partial x} = \frac{\partial x'}{\partial x}\frac{\partial }{\partial x'} + \frac{\partial y'}{\partial x}\frac{\partial }{\partial y'} +  \frac{\partial z'}{\partial x}\frac{\partial }{\partial z'} + \frac{\partial t'}{\partial x}\frac{\partial }{\partial t'}=
\frac{\partial}{\partial x'}+\frac{\gamma-1}{V^2}V_x(\vec{V} \cdot \vec{\nabla}') - \gamma \frac{V_x}{c^2} \frac{\partial}{\partial t'}
\label{PX}
\end{equation}
and similarly for the remaining 
  components:
\begin{equation}
\frac{\partial }{\partial y} =\frac{\partial}{\partial y'}+ \frac{\gamma-1}{V^2}V_y(\vec{V} \cdot \vec{\nabla}')- \gamma \frac{V_y}{c^2} \frac{\partial}{\partial t'},
\label{PY}
\end{equation}
\begin{equation}
\frac{\partial }{\partial z} =\frac{\partial}{\partial z'} +\frac{\gamma-1}{V^2}V_z(\vec{V} \cdot \vec{\nabla}')- \gamma \frac{V_z}{c^2} \frac{\partial}{\partial t'}.
\label{PZ}
\end{equation}
Using the above relation one gets:
\begin{equation}
\vec{V} \cdot \vec{\nabla} =  \gamma \vec{V} \cdot \vec{\nabla}'- \gamma \beta ^2 \frac{\partial}{\partial t'}
\label{23}
\end{equation}
where $\beta=V/c$. However on the basis of Eq. (\ref{LOR2}) we have also that:
\begin{equation}
\frac{\partial}{\partial t} = \frac{\partial x'}{\partial t}\frac{\partial}{\partial x'} +\frac{\partial y'}{\partial t}\frac{\partial}{\partial y'} + \frac{\partial z'}{\partial t}\frac{\partial}{\partial z'} + \frac{\partial t'}{\partial t}\frac{\partial}{\partial t'} =
-\gamma \vec{V} \cdot \vec{\nabla}' + \gamma \frac{\partial}{\partial t'}.
\end{equation}
Hence:
\begin{equation}
\gamma \vec{V} \cdot \vec{\nabla}' = \gamma \frac{\partial}{\partial t'} - \frac{\partial}{\partial t}.
\label{VN'}
\end{equation}
Inserting this result into Eq. (\ref{23}) we find that:
\begin{equation}
\vec{V} \cdot \vec{\nabla} = \frac{1}{\gamma }\frac{\partial}{\partial t'} - \frac{\partial}{\partial t}.
\label{24}
\end{equation}
Coming back to Eq. (\ref{22}) we obtain:
\begin{equation}
\vec{\nabla} \times \vec{B}= \frac{1}{c^2}\vec{V}(\vec{\nabla} \cdot \vec{E}) + 
\frac{1}{c^2} \frac{\partial \vec{E}}{\partial t} - \frac{1}{\gamma c^2} \frac{\partial \vec{E}}{\partial t'}.
\label{25}
\end{equation}
Using the definition (\ref{E}) and remembering that $\vec{E}'$ does not depend on time $t'$ we conclude that the last term is equal to zero. Recalling also the earlier derived  Maxwell equation (\ref{I}) and noting that $\vec{V}\rho$ is the current density $\vec{j}$ we finally arrive at the subsequent Maxwell equation:
\begin{equation}
\vec{\nabla} \times \vec{B} - \frac{1}{c^2} \frac{\partial \vec{E}}{\partial t} 
= \frac{\vec{j}}{c^2 \epsilon_0}.
\label{26}
\end{equation}\\

\section*{5. FRADAY'S LAW: \quad $ \vec{\nabla} \times \vec{E} + \frac{\partial \vec{B}}{\partial t} =0$}
Using Eqs (\ref{PX})-(\ref{PZ}) it is easy to show that:
\begin{equation}
\vec{\nabla} \times \vec{E}= \frac{\gamma-1}{V^2}\vec{V} \cdot \vec{\nabla}'(\vec{V} \times \vec{E}) + \vec{\nabla}' \times \vec{E} - \frac{\gamma}{c^2} \frac{\partial}{\partial t'}(\vec{V} \times \vec{E}).
\label{31}
\end{equation}
But recalling Eq.(\ref{VN'}) we have from (\ref{31}):
\begin{equation}
\vec{\nabla} \times \vec{E}=- \frac{\gamma-1}{\gamma V^2}\frac{\partial}{\partial t}(\vec{V} \times \vec{E}) + \vec{\nabla}' \times \vec{E} + \left(1-\frac{\gamma}{c^2} \right)\frac{\partial}{\partial t'}(\vec{V} \times \vec{E}).
\label{32'}
\end{equation}
The last term is zero because $\vec{E}$ does not depend explicitly on $t'$. So:
\begin{equation}
\vec{\nabla} \times \vec{E}=- \frac{\gamma-1}{\gamma V^2}\frac{\partial}{\partial t}(\vec{V} \times \vec{E}) + \vec{\nabla}' \times \vec{E}.
\label{32}
\end{equation}
Let us now calculate  $\vec{\nabla}' \times \vec{E}$. Firstly, from the definition (\ref{E}) follows that:
\begin{equation}
\vec{E}= \vec{E}' + (\gamma - 1) \vec{E}_\perp '.
\end{equation}
Also we can write that:
\begin{equation}
\vec{E}_\perp ' = \frac{1}{V^2}(\vec{V} \times \vec{E}') \times \vec{V}
\end{equation}
So we have:
\begin{equation}
\vec{\nabla}' \times \vec{E}= \vec{\nabla}' \times \vec{E}' + \frac{\gamma -1}{V^2}\vec{\nabla}' \times [(\vec{V} \times \vec{E}') \times \vec{V}].
\end{equation}
The field $\vec{E}'$ is an electrostatic field determined by the Coulomb law for which we have: $\vec{\nabla}' \times \vec{E}'=0$. In turn, we can 
rewrite the remaining term  using identity (\ref{ATB}). Thus the last equation is:
\begin{eqnarray}
\vec{\nabla}' \times \vec{E}= \frac{\gamma -1}{V^2} [ ( \vec{\nabla}' \times \vec{E}' )(\vec{\nabla}' \cdot \vec{V})   - \vec{V} (\vec{\nabla}' \cdot (\vec{V} \times \vec{E}'))+ \nonumber \\
+(\vec{V}\cdot\vec{\nabla}')  (\vec{V} \times \vec{E}') - (\vec{V} \times \vec{E}') \cdot\vec{\nabla}') \vec{V}].
\end{eqnarray} 
The first term in brackets is zero and, because of constancy of the "field" $\vec{V}$, also the last term vanish. It  turns out also that the second term is zero since from the well know vector identity we have:
\begin{equation}
\vec{\nabla}' \cdot (\vec{V} \times \vec{E}')= \vec{E}' \cdot (\vec{\nabla}' \times \vec{V}) - \vec{V} \cdot (\vec{\nabla}' \times \vec{E}') = 0.
\end{equation}
Thus there remains only the third term which we can evaluate using  Eq. (\ref{VN'}). As the result we get:
\begin{equation}
\vec{\nabla}' \times \vec{E}= \frac{\gamma -1}{V^2} \left( \frac{\partial}{\partial t'}
 - \frac{1}{\gamma}\frac{\partial}{\partial t}\right)(\vec{V} \times \vec{E}')= 
- \frac{\gamma -1}{\gamma V^2} \frac{\partial}{\partial t}(\vec{V} \times \vec{E}')
\label{32''}
\end{equation}
because the differentiation over $t'$ gives zero.
But:
\begin{equation}
\vec{V} \times \vec{E}'=\vec{V} \times \vec{E}_\perp'=\vec{V} \times \frac{\vec{E}_\perp}{\gamma}=\frac{1}{\gamma}\vec{V} \times \vec{E}.
\end{equation}
So Eq. (\ref{32''}) may be written as:
\begin{equation}
\vec{\nabla}' \times \vec{E}=- \frac{\gamma -1}{\gamma^2 V^2}  \frac{\partial}{\partial t}(\vec{V} \times \vec{E}).
\end{equation}
Inserting the last result into Eq. (\ref{32}) we find that:
\begin{equation}
\vec{\nabla} \times \vec{E}=- \left(\frac{\gamma-1}{\gamma V^2} + \frac{\gamma -1}{\gamma^2 V^2}  \right)  \frac{\partial}{\partial t}(\vec{V} \times \vec{E}) = -\frac{1}{c^2}\frac{\partial}{\partial t}(\vec{V} \times \vec{E})
\label{33}
\end{equation}
Recalling the definition (\ref{B}) we see that Eq. (\ref{33}) represents the Maxwell equation:
\begin{equation}
\vec{\nabla} \times \vec{E} + \frac{\partial \vec{B}}{\partial t} =0
\label{III}
\end{equation}\\

\section*{6. GAUSS' LAW FOR MAGNETISM: $\vec{\nabla} \cdot \vec{B} =0$}
Now it is easy to prove the remaining Maxwell equation. Namely,
\begin{equation}
\vec{\nabla} \cdot \vec{B} = \frac{1}{c^2} \vec{\nabla} \cdot (\vec{V} \times \vec{E})= \frac{1}{c^2} \vec{E} \cdot (\vec{\nabla} \times \vec{V}) - \frac{1}{c^2} \vec{V} \cdot (\vec{\nabla} \times \vec{E}).
\end{equation}
The first term is zero and with help of Eq. (\ref{III}) we get:
\begin{equation}
\vec{\nabla} \cdot \vec{B} =  \frac{1}{c^4} \vec{V} \cdot \left(\frac{\partial }{\partial t}(\vec{V} \times \vec{E})\right)= \frac{1}{c^4}\frac{\partial }{\partial t}  \vec{V} \cdot (\vec{V} \times \vec{E}) .
\end{equation}
It is evident that $\vec{V} \cdot (\vec{V} \times \vec{E}) $ is zero, so:
\begin{equation}
\vec{\nabla} \cdot \vec{B} = 0,
\label{IV}
\end{equation}
which ends our job.\\

\section*{7. DISCUSSION}
1. Our definition of magnetic field $\vec{B}$ (Eq. (\ref{B})) seems to differ from that one gets as a solution of Maxwell's equations:
\begin{equation}
\vec{B}=\frac{1}{c}\vec{n}_{ret} \times \vec{E}
\label{n}
\end{equation}
However, it is well known that  the solution for the field $\vec{E}$ in case of constant velocity of source $\vec{V}$ is a vector proportional to the
 instantaneous radius vector $\vec{R} \equiv \vec{r}-\vec{r}_Q$. This vector is connected with the retarded vector radius
 $\vec{R}_{ret} \equiv \vec{r}-\vec{r}_{Q_{ret}}$ by the equality $\vec{R}= \vec{R}_{ret} +\vec{V}R_{ret}/c$. It follows that
 $\vec{V}=(\vec{R}_{ret}-\vec{R})c/R_{ret}$. Inserting this formula into the definition given by Eq. (\ref{B}) we arrive at the 
equality (\ref{n}), where $\vec{n}_{ret} \equiv \vec{R}_{ret}/R_{ret}$.

2. Our reasoning is performed for the special case of source-charges moving with \emph{constant} velocity. We know that besides the fields determined by such sources Maxwell's equations offer us as their general solutions fields that additionally depend on the acceleration of source-charges. This may be regarded as an unexpected auxiliary discovery suggested by our reasoning. 
We emphasize that it does \emph{not} follow from our reasoning that the general solutions are physically correct because we have neither defined the fields $\vec{E}$ and $\vec{B}$ for accelerating sources nor have proved that in this case the force have the form of the Lorentz force. Existence of such fields and forces is merely suggested by the derived results.
It means that  one must verify this fact \emph{experimentally}. Forces produced by accelerating charges should agree with the Lorentz force law (\ref{EB1}) in which inserted are the general solutions of Maxwell's equations. So far all the experimental data confirm that the Lorentz force law and the discovered here theoretically general solutions of Maxwell's equations are valid.

To avoid confusion let us mention that the relations between electromagnetic fields in different systems of reference given in Eqs. (\ref{E}) 
and (\ref{B}) do \emph{not} apply to the general case of accelerating sources. The reason is that in case of non-uniformly moving 
charges at a space-time point $(\vec{x}', t')$ in the rest frame of the source, if only at the retarded moment the source accelerated, there exist a magnetic field 
$\vec{B}'$. The force in such a frame is not then the Coulomb one but Lorentzian force $ \vec{F}'= q\vec{{E}}' + q\vec{v}' \times \vec{{B}}'$.  
The transformation (\ref{FFF}) allows us to find the correct relations between the fields ($\vec{E}$, $\vec{B}$) and ($\vec{E}'$, $\vec{B}'$):
\begin{equation}
\vec{E}=\vec{E}'_\parallel +\gamma \vec{E}'_\perp -\gamma\vec{V} \times \vec{B}',
\label{E''}
\end{equation}
\begin{equation}
\vec{B}=\vec{B}'_\parallel +\gamma\vec{B}'_\perp +\frac{\gamma}{c^2} \vec{V} \times \vec{E}'.
\label{B''}
\end{equation}
Eqs. (\ref{E}) 
and (\ref{B}) are then only a special case of the above ones (for the field $\vec{B}'$ equal to zero).

Similarly, the reader should be aware that the correct \emph{general} relation between the fields $\vec{E}$ and $\vec{B}$  is not (\ref{B}) but (\ref{n}). 

3. We have derived Maxwell's equations for a single point charge. Because these equations are linear and thanks to the Principle of Superposition the same equations are correct for electromagnetic fields produced by any distribution of many charges.

4. One may be tempted to apply our method to the gravitational Newton force which formally may seem to be  identical to Coulomb's force and obtain 
Maxwell's equation for the gravitational interactions. This procedure would however be wrong because if one wants to consider the gravitational force in
 the realm of relativity (without using the General Theory of Relativity)  to get correct results it must be assumed that the gravitational mass depends on 
the velocity of moving body. It is not the case if concerns the charge $Q$ which is velocity-independent.  The method presented in this work applies then
 only to the electromagnetic forces.\\

\section*{8. CONCLUSIONS}
Coulomb's law together with the Lorentz transformation of force have led us to electromagnetic fields produced by uniformly moving sources. 
These fields have appeared to be governed by Maxwell's equations. In turn, if Maxwell's equations are solved, they deliver us a wider class 
of solutions that could apply to arbitrarily moving sources. Experiments confirm this unexpected prediction.  In this way the reasoning presented 
in this paper  is an example of a scientific discovery based solely on the theoretical argumentation. However, as for any kind of theoretical results,
 their experimental verification  is indispensable. \\

\section*{Acknowledgments}
I would like specially to thank P. Czas who inspired me to write this paper. I am grateful to Polish Ministry of Science and Information Society 
Technologies (State Committee for Scientific Research) for financial support of this work.\\

\section*{APPENDIX}
Let a reference system $S'$ moves from the point of view of some other system $S$ with constant velocity $\vec{V}$.
In the reference system $S'$ the equation of motion is:
\begin{equation}
\vec{F}'=\frac{d\vec{p}\;'}{dt'}
\label{S'}
\end{equation}
and in the system $S$:
\begin{equation}
\vec{F}=\frac{d\vec{p}}{dt}.
\end{equation}
Each of these equations may be decomposed  on the components along and the directions parallel and perpendicular to the velocity $\vec{V}$:

\begin{equation}
\vec{F}'_\parallel=\frac{d\vec{p}\;'_\parallel}{dt'}, \qquad \vec{F}'_\perp=\frac{d\vec{p}\;'_\perp}{dt'}
\end{equation}
and
\begin{equation}
\vec{F}_\parallel=\frac{d\vec{p}_\parallel}{dt}, \qquad \vec{F}_\perp=\frac{d\vec{p}_\perp}{dt}.
\end{equation}
To find the relation between forces $\vec{F}$ and $\vec{F}'$ let us recall the Lorentz transformation for momentum and time:

\begin{equation}
\begin{array}{l@{\vspace{0.3 cm}}}
\vec{p}\;'_\parallel=\gamma\left(\vec{p}_\parallel -E\frac{\vec{V}}{c^2}\right),\\
\vec{p}\;'_\perp=\vec{p}_\perp
\label{P}
\end{array}
\end{equation}
and
\begin{equation}
t'=\gamma\left(t-\frac{\vec{r}\cdot \vec{V}}{c^2}\right),
\end{equation}
where $\gamma=(1- V^2/c^2)^{-1/2}$. We have then that:
\begin{equation}
\frac{dt}{dt'}= \frac{1}{\gamma\left(1-{\vec{v}\cdot \vec{V}}/{c^2}\right)},
\label{DT}
\end{equation}
where $\vec{v}=d\vec{r}/dt$ is a velocity of body measured in the frame $S$.
First, using the second equation of (\ref{P}) and Eq. (\ref{DT}), we find that:
\begin{equation}
\vec{F}'_\perp=\frac{dt}{dt'}\frac{d\vec{p}_\perp}{dt}=
\frac{1}{\gamma\left(1-{\vec{v}\cdot \vec{V}}/{c^2}\right)} \vec{F}_\perp.
\label{FP}
\end{equation}
In turn, for the parallel component of force we obtain from the first equation of  (\ref{P}) and Eq. (\ref{DT}) that:
\begin{equation}
\vec{F}'_\parallel=\gamma\frac{dt}{dt'}\frac{d}{dt}\left(\vec{p}_\parallel -E\frac{\vec{V}}{c^2}\right)=
 \frac{1}{1-{\vec{v}\cdot \vec{V}}/{c^2}} \left(\vec{F}_\parallel -\vec{F} \cdot \vec{v} \frac{\vec{V}}{c^2}\right),
\label{FR1}
\end{equation}
where we have substituted $dE/dt=\vec{F} \cdot \vec{v}$.
Now using the well known vector identity:
\begin{equation}
\vec{A} \times (\vec{B} \times \vec{C})= (\vec{A} \cdot \vec{C})\vec{B}- (\vec{A} \cdot \vec{B})\vec{C}
\end{equation}
we get:
\begin{equation}
\vec{F} \cdot \vec{v} \frac{\vec{V}}{c^2}= \vec{v} \times \left({\frac{\vec{V}}{c^2}} \times \vec{F}\right)+\left(\frac{\vec{v} \cdot \vec{V}}{c^2}\right)\vec{F}.
\end{equation}
Thus Eq. (\ref{FR1}) may be written as:
\begin{equation}
\vec{F}'_\parallel= \frac{1}{ 1-{\vec{v}\cdot \vec{V}}/{c^2}} \left(\vec{F}_\parallel - \frac{ \vec{v} \cdot \vec{V}}{c^2}\vec{F}_\parallel-
\frac{\vec{v} \cdot \vec{V}}{c^2}\vec{F}_\perp- \vec{v} \times \left({\frac{\vec{V}}{c^2}} \times \vec{F} \right)\right),
\end{equation}
or in a simpler form:
\begin{equation}
\vec{F}'_\parallel=\vec{F}_\parallel - \gamma\frac{\vec{v} \cdot \vec{V}}{c^2}\vec{F}'_\perp- \frac{\vec{v} \times \left({\frac{\vec{V}}{c^2}} \times \vec{F}\right)}{1-{\vec{v}\cdot \vec{V}}/{c^2}}.
\label{FR2}
\end{equation}
However, using Eq. (\ref{FP}) we get:
\begin{equation}
\vec{V} \times \vec{F}=\vec{V} \times \vec{F}_\perp= \gamma \left(1-\frac{\vec{v}\cdot \vec{V}}{c^2}\right) \vec{V} \times \vec{F}'_\perp= 
 \gamma \left(1-\frac{\vec{v}\cdot \vec{V}}{c^2}\right) \vec{V} \times \vec{F}'.
\end{equation}
Because of this the third term on the right hand side of Eq. (\ref{FR2}) can be much simplified and we have:
\begin{equation}
\vec{F}'_\parallel=\vec{F}_\parallel - \gamma\frac{\vec{v} \cdot \vec{V}}{c^2}\vec{F}'_\perp- \gamma\vec{v} \times \left({\frac{\vec{V}}{c^2}} \times \vec{F}' \right).
\label{FR3}
\end{equation}
Finally then from the last equation we obtain:

\begin{equation}
\vec{F}_\parallel=\vec{F}'_\parallel + \gamma\frac{\vec{v} \cdot \vec{V}}{c^2}\vec{F}'_\perp+ 
\gamma\vec{v} \times \left({\frac{\vec{V}}{c^2}} \times \vec{F}' \right)
\label{FR4}
\end{equation}
and from Eq. (\ref{FP}):

\begin{equation}
\vec{F}_\perp={\gamma\left(1-\frac{\vec{v}\cdot \vec{V}}{c^2}\right)} \vec{F}'_\perp.
\label{FP1}
\end{equation}
Adding up the last two equations we find that:
\begin{equation}
\vec{F}=\vec{F}'_\parallel + \gamma\vec{F}'_\perp+ \gamma\vec{v} \times \left(\vec{\frac{V}{c^2}} \times \vec{F}' \right),
\label{FF}
\end{equation}
which is the desired relation between the forces $\vec{F}$ and $\vec{F}'$.\\

\section*{REFERENCES}
\begin{enumerate}
\item R. P. Feynman, R. B. Leighton, M. Sands, \emph{The Feynman Lectures on Physics} (Addison-Wesley, Massachusetts, 1963), Sec. 26-1.
\item W. G. V. Rosser, \emph{Classical Electromagnetism via Relativity} (Butterworths, London, 1968).
\item D. H. Frish, L. Wilets, "Development of the Maxwell-Lorentz Equations from Special Relativity and Gauss's Law," 
Am. J. Phys. \textbf{24}, 574-579 (1956).
\item R. H. Lehmberg, "Axiomatic Development of the Laws of Vacuum Electrodynamics,"
Am. J. Phys. \textbf{29}, 584-592 (1961).
\item E. Krefetz, "A 'Derivation' of Maxwell's Equations," Am. J. Phys. \textbf{38}, 513-516 (1970).
\item D. H. Kobe, "Generalization of Coulomb's Law to Maxwell's equations using Special
Relativity," Am. J. Phys. \textbf{54}, 631-636 (1986).
\item J. R. Tessman, "Maxwell - Out of Newton, Coulomb, and Einstein," Am. J. Phys. \textbf{34}, 1048-1055 (1966). 
\item E.M. Purcell, \emph{Berkeley Physics Course}, Vol.2 (McGraw-Hill, New York, 1963), Sec. 5.6-5.9.
\item M. Schwartz, \emph{Principles of Electrodynamics} (McGraw-Hill, New York, 1972), Sec. 3.
\item J. D. Jackson, \emph{Classical Electrodynamics}  (Willey, New York, 1999), 3rd ed., Sec. 11.3.
\end{enumerate}
\end{document}